# Characterization of SOI pixel sensor using pinned depleted diode structure with AC-coupling on the diode

Jing Dong [a, 1], Yunpeng Lu [a, 1], Zhigang Wu [a, 2], Yang Zhou [a], Qun Ouyang [a, b]

[a] *State Key Laboratory of Particle Detection and Electronics (Institute of High Energy Physics, CAS), Beijing 100049, China*
[b] *University of Chinese Academy of Sciences, Beijing 100049, China*

Jing Dong: dongj@ihep.ac.cn
Yunpeng Lu: yplu@ihep.ac.cn

**Abstract:** The experiment of the future electron-positron colliders has unprecedented requirements on the vertex resolution, such as around 3μm single point resolution for the inner most detector layer, with fast readout, and very low power-consumption density and material budget. Significant efforts have been put into the development of monolithic silicon pixel sensors, but there have been some challenges to combine all those stringent specifications in a small pixel area. This paper presents a compact prototype pixel sensor fabricated in LAPIS 200nm SOI process and focuses on the characterization of low capacitance of the sensing node with a pinned depleted diode (PDD) structure adopting a novel method of forward bias voltage and AC-coupling on the diode. Three PDD structures with 16×20μm$^2$ pixel size were designed and compared using radioactive sources and injected charge. The measured result shows that the designed PDD structure has very low leakage current and around 3.5fF of equivalent input capacitance.

Contents

## 1  Introduction

The experiment on the next generation of electron-positron collider as a Higgs factory should have to reach the unprecedented impact parameter resolution, for the purpose of efficient tagging of heavy flavor quarks and τ lepton. This sets stringent demands on silicon pixel sensors of the vertex system, which should be developed with small pixel size (~16 μm) and low noise front-end. The silicon-on-insulator (SOI) technology [1], featuring as a monolithic silicon pixel process with fully depleted substrate, has potential to meet those basic requirements. We have been developing an SOI pixel sensor prototype using Double-SOI process with the pixel size of $16\times16$ μm$^2$, which can achieve single point resolution below 3 μm with an infrared laser beam [2]. Since 2017, a new generation sensor process called pinned depleted diode (PDD) has been developed, which has shown its features [3, 4, 5] such as high charge collection efficiency, suppressing the leakage current at the Si-SiO2 interface, minimizing the junction capacitance and so on. In order to take advantages of PDD in our **c**ompact **p**ixel detector series chips for **v**ertex (CPV) and study their performances such as diode capacitances, equivalent input capacitance, and charge sharing and so on, three different PDD structures plus one null structure were developed to learn the characterization of the different structures in the current chip version CPV3.

This paper is dedicated to the evaluation of equivalent input capacitance of CPV3, which has been fabricated in the LAPIS 200nm SOI process. The description of the prototype chip including PDD sensor diode, CPV3-PDD structures，pixel schematic and test setup is presented in Section 2, followed by the characterization results of the sensors in Section 3. Finally in Section 4, some conclusions will be drawn and future plans will be outlined.

## 2  Description of the prototype chip

The prototype chip CPV3 has an area of $6 \times 6$ mm$^2$, which includes a sensitive pixel matrix of $4 \times 4$ mm$^2$. The pixel array was divided into four sensor matrices in the view of row, and five main circuit matrices in the view of column for analog or digital readout. The part we are studying now focusing on PDD sensor matrices with step injected charge structure and analog readout, is on the left part of the chip consists of $204 \times 40$ identical cells with a pixel size of $16 \times 20$ μm$^2$. Except the second PDD structure matric that consists of $108 \times 40$ identical cells, all the other two PDD matrices plus one null structure are consisting of $32 \times 40$ identical cells. In-pixel signal processing and peripheral bias and controls are on the thin top layer of the sensor which is a few microns thickness.

### 2.1  PDD

CPV3 was designed with PDD whose basic schematic diagram is shown in Figure 1. It is mainly composed of several layers of doped structures with different depths from the Si and SiO2 interface. High negative voltage such as minus sixty volts is applied at the backside of the sensor for obtaining a fully depleted thick substrate. NS and BNW1 form a charge collector, which is connected to the metal layer. The highly doped BPW1 layer acts as a shielding layer between the circuits [6] and the charge collector and is connected to a fixed potential to prevent crosstalk between the circuit layer and the sensor. Meanwhile the BPW1 layer blocks the contact between the sensitive area of the sensor and the Si-SiO2 interface, which could reduce the dark current from the interface effectively, and improve the charge collection efficiency. By adding a certain negative bias to BPW1 and BNW1 layers, both layers of BNW2 and BNW3 get fully depleted, which makes the effective area of charge collector greatly shrink, therefore reducing the capacitance of the sensor. Moreover, the multi-level doped structure will form a lateral gradient electrical field, which is beneficial to the charge collection efficiency.

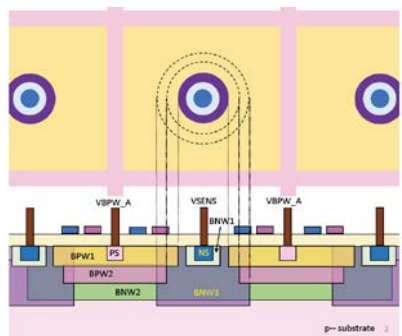

**Fig. 1.** The schematic view of the PDD diode structure.

## 2.2 CPV3-PDD structure

We have designed three different sensor-structures and one null structure in order to learn the characterization in such small pixel dimension. The first type CPV3-PDD1 is shown in Figure 2(left). The charge collector adopts the smallest regular octagon (diameter 2.8μm), and the distance between BNW1 and BPW1 is 2μm. Counting from the center to the outside, the radius is: NS/0.6μm, BNW1/1.4μm, BPW1/3.4μm, BPW2/4.9μm, BNW3/6.4μm, respectively. BNW2 covers the entire pixel area.

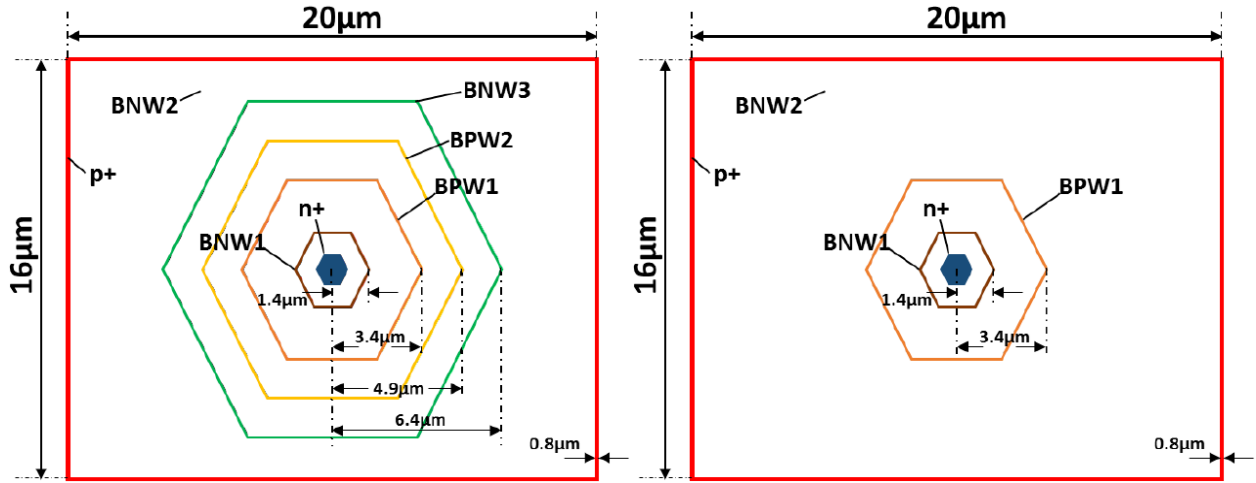

**Fig. 2.** The CPV3-PDD1 structure (left) and simplified PDD (CPV3-PDD2) structure (right).

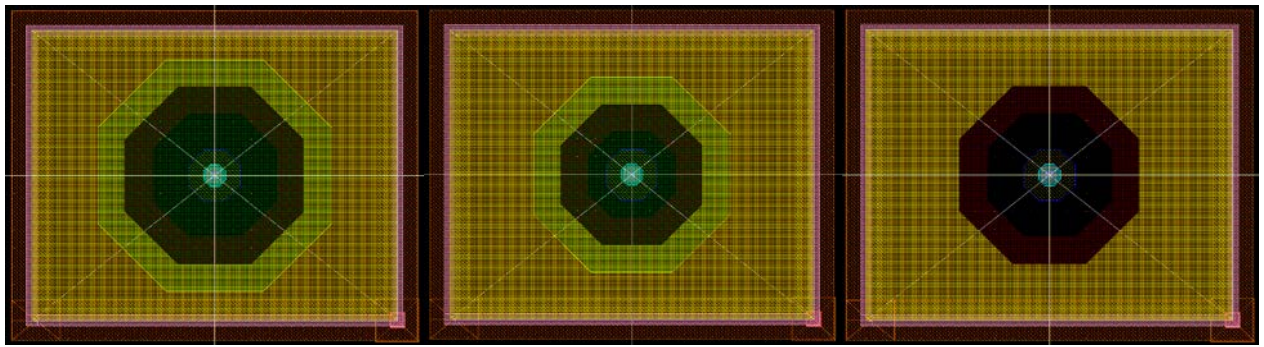

（a）. CPV3-PDD1　　　　(b). CPV3-PDD2　　　　（c）. CPV3-PDD3
**Fig. 3.** The three different sensors structures.

Table 1: The main parameters of the structures

| Unit：μm Type | Diameter of charge collector octagon | The distance between BNW1 and BPW1 | The radius counting from the center to the outside | | | | |
|---|---|---|---|---|---|---|---|
| | | | NS | BNW1 | BPW1 | BPW2 | BNW3 |
| CPV3-PDD1 | 2.8 | 2 | 0.6 | 1.4 | 3.4 | 4.9 | 6.4 |
| CPV3-PDD2 | 2.8 | 1 | 0.6 | 1.4 | 2.4 | 3.9 | 5.4 |
| CPV3-PDD3 | 2.8 | 2 | 0.6 | 1.4 | 3.4 | No | No |

The CPV3-PDD2 has a similar structure to the previous one, the main difference from it is that the distance between BNW1 and BPW1 is shortened to 1um to investigate the impact of the distance to the diode capacitance. Counting from the center to the outside, the radius is: NS/0.6μm, BNW1/1.4μm, BPW1/2.4μm, BPW2/3.9μm, BNW3/5.4μm, respectively. BNW2 covers the entire pixel area. The third type CPV3-PDD3 is a simplified PDD structure, which is shown in Figure 2(right). On the basis of CPV3-PDD1, the layers of BPW2 and BNW3 are removed to compare its effect on the edge of the pixel. The figure 3 is shown how the diodes look like and table 1 summarizes the comparison of the main parameters of the structures.

### 2.3 Pixel schematic

From the results of the TCAD simulation, at least a 4.4V bias voltage must be applied between BNW1 and BPW1 to reduce the diode capacitance below 4 fF. For the SOI-PDD sensor, the traditional method is to use 3.3V PMOS as the input, so that a +2 V voltage can be provided to the charge collector, and a -2 V voltage is applied to the BPW1 layer. For CPV3, due to the limited dimension of the size, it is unacceptable for 3.3 V circuit design, which means that we need to apply at least 3-4 V negative bias on the BPW1 layer. However, the BPW1 layer is just below the circuit layer, with only a thin layer of SiO2 in between. The change of the potential of the BPW1 layer will cause the shift of the transistors' threshold in the circuit layer, which lead the circuit to fail to work normally. In order to solve this problem, a novel method that uses forward bias voltage and AC coupling on the diode has been adopted. Figure 4 shows the schematic in each pixel cell of CVP3 PDD test structure. For calibration, all the matrices are analog pixels. A pulse-test structure with a capacitor that serves the purpose of injecting a controlled step-charge were placed. By measuring the energy spectrum of $^{55}Fe$ radiation source as well as injected step-charge, the electrical characterization such as the equivalent capacitance of diode Cdiode, parasitic capacitance Cp and pulse-test capacitance Cinj could be mostly evaluated. The whole pixel array is read out using a rolling shutter mode to decrease power consumption. The common source stage is always switched off except for the time when the specific row of pixels is selected for readout.

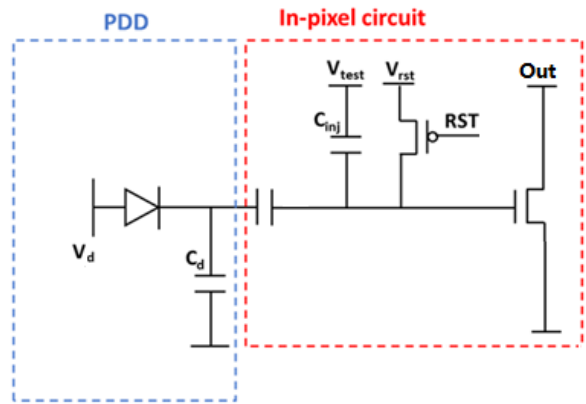

**Fig. 4.** The pixel schematic the electronics.

### 2.4 Experimental setup

The test system consists of the CPV3 chip, the data acquisition board and the DAQ software. The chip is wire-bonded on dedicated chip carrier PCB which in turn are connected to the commercial FPGA KC705 board. The timing and reference voltages are controlled by the FPGA programming on the KC705 and the DAQ software on PC. The analog signal is converted to digital readout by a 12 bits ADC which has an upper limit range of 2 V.

## 3 Characterization results

### 3.1 Leakage current

The current voltage characteristics were measured. In Figure 5(left) shows the sources of the leakage current and measurements as a function of each source have been performed. As it can be seen from fig.5 (right), the total leakage currents increase as the bias voltages increase, the number is quite low from the PDD diode which has shown a quite good performance as expected.

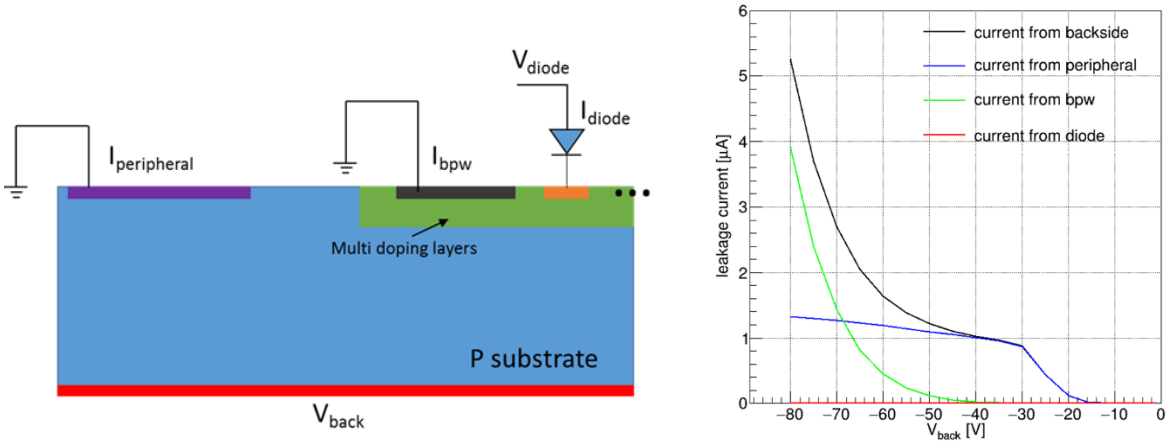

Fig. 5. The sensor structure of CPV3 (left) and the total leakage current as a function of the bias voltage (right).

### 3.2 Gain

The circuit gain's measurements of three PDD structures have been performed. In Figure 6, the output of the whole pixel array of CPV3-PDD1 is presented as a function of the voltage Vrst. The slope of the curve in the left figure is the gain of the circuit which has an average gain equal to 0.87. The right figure shows the distribution of the gain with the RMS value of the gain distribution equal to 0.004. The other two structures show the quite similar performances.

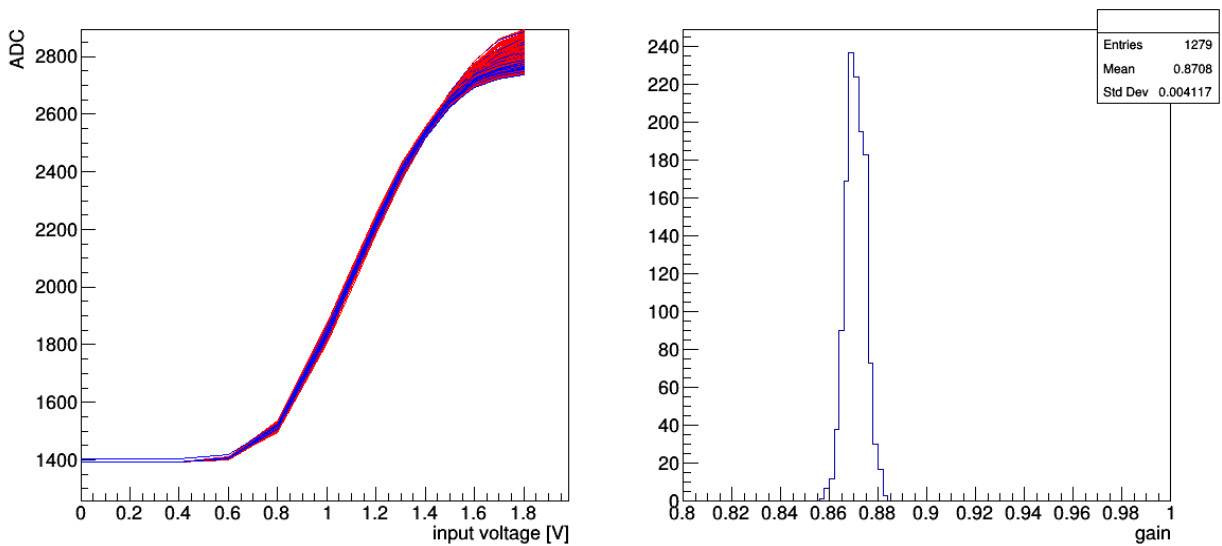

Fig. 6. The analog readout as a function of the voltage Vrst for the whole pixel array. (a)The slope of the curve is the gain. (b) The distribution of the gain.

### 3.3 Calibration

#### 3.3.1 The energy spectra of $^{55}$Fe

$^{55}$Fe X-ray source was used to measure the charge voltage factor (CVF) and the equivalent input capacitance that equal to the sum of diode Cdiode and the parasitic capacitance Cp in the PDD structures. Fig. 7 shows the energy spectra of CPV3-PDD1 in ADC units using two modes. They are the charge collected on a matrix signal composed of 3 × 3 pixels and 5 × 5 pixels respectively with the threshold of the seed as 10sigma. In the two matrix signals, a peak at 38.58 and 38.63 ADC can be observed and the matrix 3 × 3 pixels is enough to collect the whole charge. In the cluster finding algorithm, the pixel with the largest analog readout is chosen as the seed pixel. The matrix signal consisting of the seed pixel in the center and the other closest neighbor pixels, is the sum of the signals of an n × n matrix. The threshold is set to avoid the noise counts. Any pixel with the analog readout below the threshold will not be taken as the seed pixel. Ten sigma as a threshold

of the seed has been learned in order to reach the entry of the noise signal equal to zero. There are two reasons we use the matrix signal, one is the probability for CPV3 chip that the charge responds only to a single pixel is very few. The other is that if we are looking for the cluster consisting with several pixels, we need to set the thresholds for all the members of the cluster, it will strongly depend on the thresholds we set. Therefore, it's more reasonable in the case of an n × n matrix signal with the condition of setting the seed' threshold only. Therefore, for CPV3-PDD1, the peak at 38.6 ADC corresponds to the 5.9 keV. Assuming 3.6 eV is needed to produce an electron–hole pair in silicon, the CVF and input capacitance can be calculated as:

$$\text{CVF} = \frac{V}{\text{Nelectron}} = \frac{38.6 \times 2/(4095 \times 0.87)}{5900/3.6} = 13 \ \mu\text{V/e}-,$$

$$\text{C}_{\text{input}} \text{(CPV-PDD1)} = \frac{Q}{V} = \frac{\left(\frac{5900}{3.6}\right) \times 1.6 \ast 10^{-19}}{38.6 \times 2/(4095 \times 0.87)} = 12\text{fF}.$$

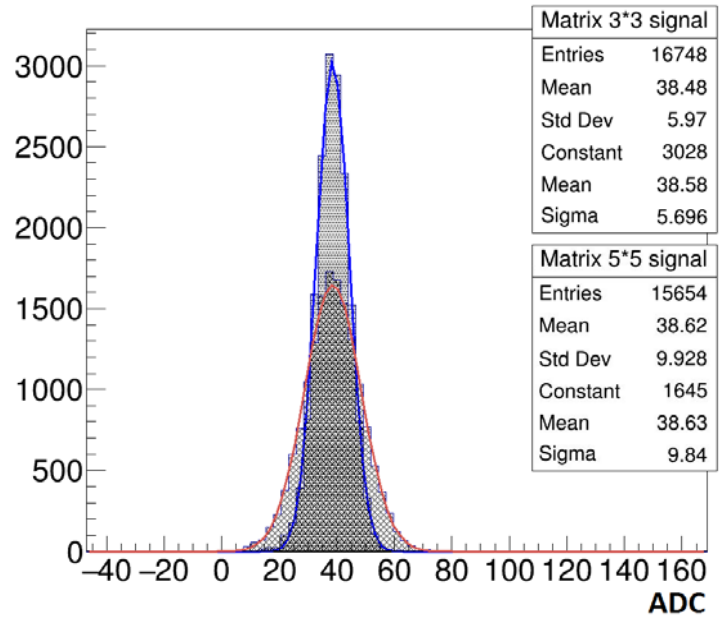

Fig. 7. The energy spectra of matrix 3×3, 5×5 signal of CPV-PDD1.

Certainly as a prerequisite, each single channel pixel' output waveform from the raw data was carefully checked with (or not) $^{55}$Fe x-ray radiation source. As presented in figure 8, the readout of a random pixel without x-ray has shown that it has a pedestal around 2298 ADC units and a noise around 5 ADC units. (Figure 9) The signal which equal to the subtraction of the readout data from its own average value of pedestal, has shown a good proportion of signal/noise. The reason that readout value is less than pedestal when signal happens, is due to the ADC module' mechanism we selected.

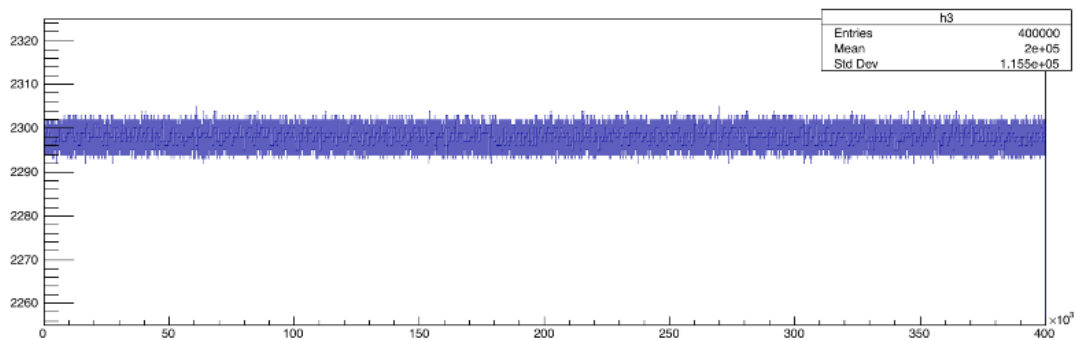

Fig. 8. The waveform of a single channel without $^{55}$Fe radiation source

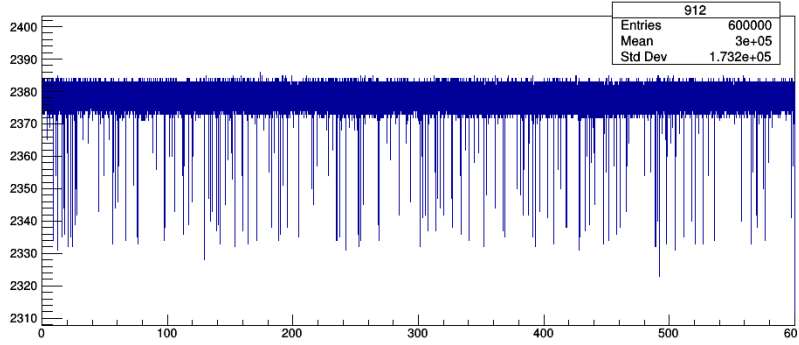

Fig. 9. The waveform of a single channel with $^{55}$Fe radiation source

### 3.3.2 Capacitance measurements

The pulse test capacitor Cinj as a function of bias voltages have been performed through the CPV3-PDD1 array via the controlled step injected charge with each measured value of equivalent input capacitance by x-ray corresponding to the bias voltages. In Figure 10(left), capacitance Cinj is presented as a function of the bias voltage; the measured mean value is 0.667fF. In addition, we repeat the same measurement but through the null array where the input capacitance is the parasitic capacitance Cp indeed. The parasitic capacitance was evaluated as shown in fig. 10(right).

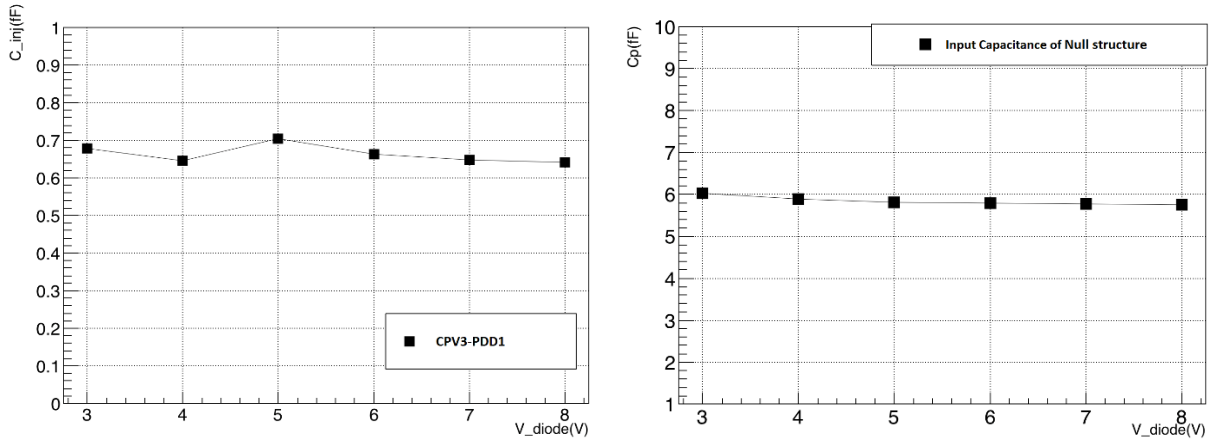

Fig.10.Capacitance of step-charge Cinj (left) and the parasitic capacitance (right) as a function of bias voltage.

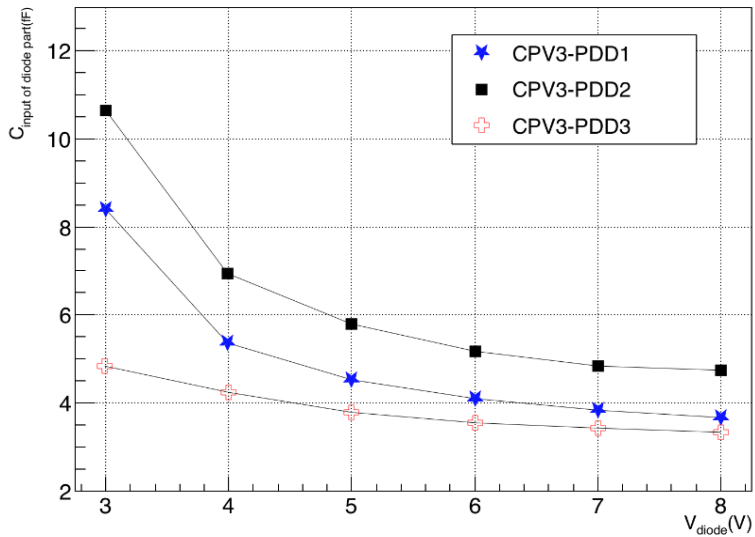

Fig.11.The equivalent diode capacitance as a function of the forward bias voltage.

Thanks to the parameters Cp and Cinj are constant approximately, the comparison of the three PDD sensors was performed via pulse injected charge. The results in Fig.11 shows the comparison of the equivalent diode capacitance which decreases as the forward bias voltage increases and CPV3-PDD3's is the lowest.

### 3.3.3 The cluster multiplicity

Charge sharing was always taken advantages in the spatial resolution' study, for our three CPV3-PDD structures, the cluster multiplicity has been carefully taken into account due to the thickness of the chip and PDD gradient electric field which will minimize the charge sharing. The cluster multiplicity is defined as the number of pixels assigned to a cluster. In the design of CPV3 chips, we estimate this number will be as maximum as 4 pixels, and higher number will be advantageous for improving the resolution. The goal is to find the specific structure with such kind of cluster multiplicity, and it's the reason we designed a simplified structure for CPV3-PDD3 in order to check the last layers' effect on the electric field. In the section 3.31, we have explained the way to set threshold of the seed, in addition, for the neighbors close to the seed, it can be decided by using the rule of 'Fake Hit Rate' when the noise entries accounted for a certain small percentage （Entries/real data counts<$10^{-6}$）. With this mechanism, thresholds of CPV3-PDD1/2/3 are set as Seed/Neighbor = 10/5sigma, Seed/Neighbor = 15/10sigma, Seed/Neighbor = 6/3.8sigma, respectively. Fig.12 (a/b/c) shows the cluster multiplicity distribution of the three PDD structures, it's 1.37, 1.01 and 1.55. It confirms our estimation that it is 4 pixels maximally when it happens closed to the cross center and the beneficial multiplicity is from the CPV3-PDD3 structure.

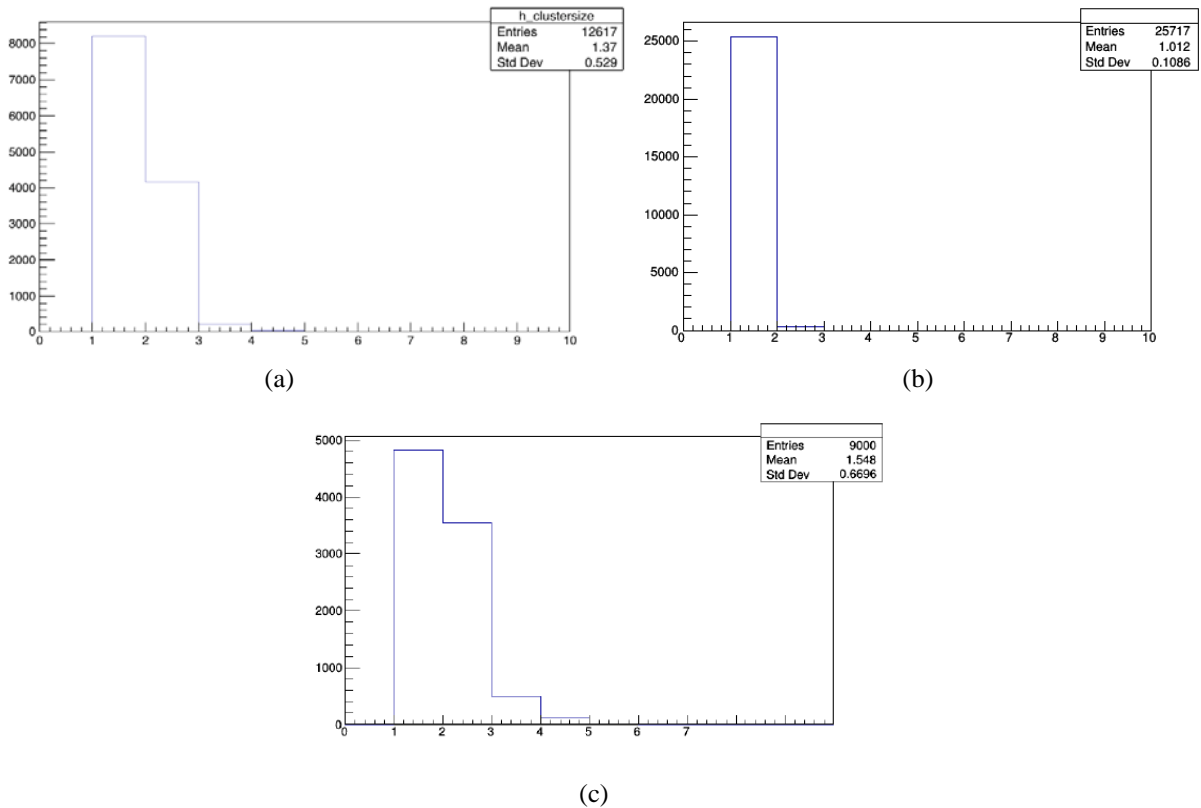

Fig.12. Cluster multiplicity distribution of CPV3-PDD. (a) CPV3-PDD1 (Seed>10sigma, neighbor>5sigma). (b) CPV3-PDD2 (Seed>15sigma, neighbor>10sigma). (c) CPV3-PDD3 (Seed>6sigma, neighbor>3.8sigma).

Since in the cluster finding algorithm, it is strongly dependent on the choice of the thresholds used for the assignment of seed and neighbor. In order to compare them under the same condition, after compromising the noise level of the three, we set the same thresholds for all the three structures' seed and neighbor that is Seed/Neighbor = 10/5sigma. Fig.13 shows the cluster multiplicity distribution as a function of the forward bias voltage of the three PDD structure with the same choices as Seed>10sigma, Neighbor>5sigma. Although the noise level of diffident structures are not the same, CPV3-PDD3 still have shown a better assignment than the other two CPV3-PDDs experimentally.

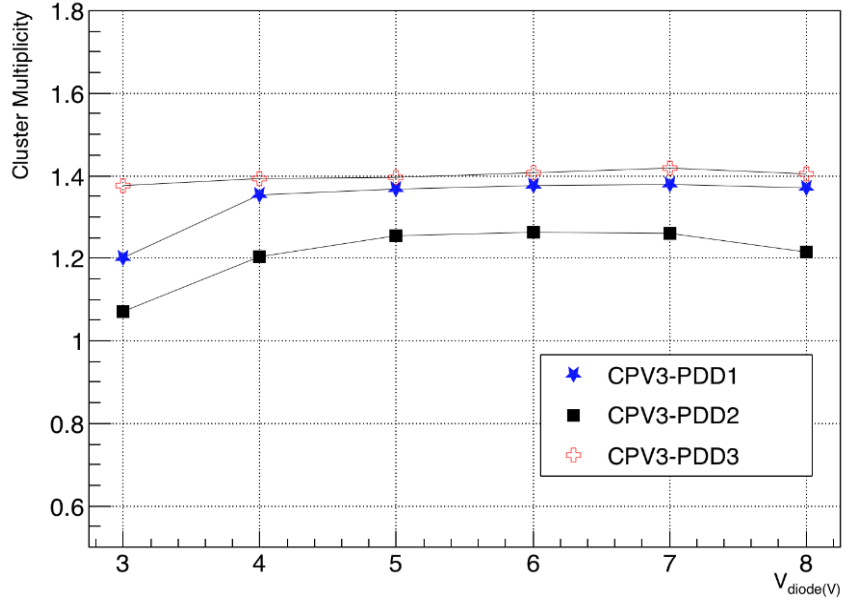

Fig.13. the cluster multiplicity distribution as a function of the forward bias voltage of the three PDD structures. (Seed>10sigma, neighbor>5sigma)

## 4  Conclusion and outlook

Three PDD sensor structures in the version CPV3 chip featuring $16 \times 20$ $\mu m^2$ pitch and 310 $\mu m$ thickness have been studied. It proposes an innovative method that uses forward bias voltage and AC coupling on the PDD structure within one pixel size. A complete overview of the characterization performed on three PDD sensors including capacitance- , current-voltage characteristics and cluster multiplicity was reported by measuring the energy spectrum of $^{55}Fe$ 5.9keV x-ray as well as pulse injected charge. The measured result shows that the designed PDD structure has very low leakage current, the equivalent input capacitance decreases as the forward bias voltage increases and around 3.5fF of equivalent in-pixel capacitance. From the comparison, we get the conclusion that one of the three structure shows lower junction capacitance and better charge sharing. The evaluation was carefully taken into account and chosen in the design of the next generation CPV4 chip with SOI-3D technology.

**Acknowledgments**

This work is supported by the National Nature Science Foundation of China, Grant 11935019.